\documentclass{iopart}

\usepackage{iopams}

\pdfoutput=1
\usepackage {amssymb}
\usepackage{setstack}

\usepackage {graphicx}
\usepackage{hyperref}
\usepackage{color}
\usepackage{tensor}

\newcommand{\B}{\mathcal{B}}
\newcommand{\F}{\mathcal{F}}

\newcommand{\Sn}{{S_\mathrm{n}}}

\newcommand{\scalar}[2]{\left(#1|#2\right)}

\newcommand{\Hyp}{\mathcal{H}}
\newcommand{\Lr}{\mathcal{L}}
\newcommand{\pdf}{\mathrm{pdf}}
\newcommand{\fA}{f_A}
\newcommand{\fD}{f_D}
\newcommand{\Om}{\Omega}
\newcommand{\vOm}{\vec{\Om}}
\newcommand{\eps}{\varepsilon}
\newcommand{\Izz}{I_{\mathrm{zz}}}
\newcommand{\cosi}{\cos\iota}
\newcommand{\cosSQi}{\cos^2\!\iota}
\newcommand{\nhat}{\hat{n}}
\newcommand{\uhat}{\hat{u}}
\newcommand{\vhat}{\hat{v}}
\newcommand{\xihat}{\hat{\xi}}
\newcommand{\zihat}{\hat{\zeta}}

\newcommand{\etens}{\tens{e}{}}

\newcommand{\htens}{\tens{h}{}}

\newcommand{\Freq}{f}
\newcommand{\fdot}{{\dot{\Freq}}}
\newcommand{\fddot}{\ddot{\Freq}}

\newcommand{\mubar}{{i}}

\newcommand{\sig}{{\mathrm{s}}}

\newcommand{\Stat}{\mathfrak{S}}

\newcommand{\Ap}{A_{+}}
\newcommand{\Ac}{A_{\times}}

\newcommand{\Dop}{\lambda}
\newcommand{\Amp}{\mathcal{A}}
\newcommand{\AmpPhys}{\bar{\Amp}}

\newcommand{\AmpSpace}{\mathbb{A}}

\newcommand{\ML}{\mathrm{ML}}

\newcommand{\M}{\mathcal{M}}

\newcommand{\thresh}[1]{{#1}_*}

\newcommand{\NMC}{N_{\mathrm{MC}}}

\newcommand{\rad}{\mathrm{rad}}

\newcommand{\Noise}{{\mathrm{N}}}
\newcommand{\Signal}{{\mathrm{S}}}

\newcommand{\Prior}{\Pi}
\newcommand{\canonical}{\mathrm{c}}
\newcommand{\physical}{\mathrm{ph}}

\newcommand{\pA}{\Prior_\AmpSpace}

\newcommand{\dcc}{LIGO-P0900066-v2}

\newcommand{\tens}[1]{\tensor{#1}}

\def\commitID{commitID: 5693fe17cf708ffc07067fc1c39814c18fba783b}
\def\commitDATE{Wed Jul 15 13:51:46 2009 +0200}

\begin{document}

\title[Bayesian versus maximum-likelihood statistics]{Targeted search for continuous gravitational waves: Bayesian versus maximum-likelihood statistics}

\author{Reinhard Prix$^1$, Badri Krishnan$^2$}
\address{$^1$ Albert-Einstein-Institut, Callinstr.\ 38, 30167 Hannover, Germany}
\address{$^2$ Albert-Einstein-Institut, Am M\"u{}hlenberg 1, 14476 Potsdam, Germany}

\ead{Reinhard.Prix@aei.mpg.de}



\address{}
\address{\dcc\qquad\commitDATE \\\mbox{\small \commitID}}

\begin{abstract}
We investigate the Bayesian framework for detection of continuous gravitational waves
(GWs) in the context of targeted searches, where the phase evolution of the GW signal is
assumed to be known, while the four amplitude parameters are unknown.
We show that the orthodox maximum-likelihood statistic (known as $\F$-statistic) can be rediscovered
as a Bayes factor with an unphysical prior in amplitude parameter space.
We introduce an alternative detection statistic (``$\B$-statistic'') using the Bayes factor with a
more natural amplitude prior, namely an isotropic probability distribution for the
\emph{orientation} of GW sources.
Monte-Carlo simulations of targeted searches show that the resulting Bayesian \mbox{$\B$-statistic} is more powerful in the
Neyman-Pearson sense (i.e.\ has a higher expected detection probability at equal false-alarm
probability) than the frequentist $\F$-statistic.
\end{abstract}



%
\section{Introduction}


Searches for gravitational waves (GWs) often consist of testing the data for the presence of signals from a
\emph{known family} of waveforms, parametrized by (generally unknown) signal parameters.
Here we consider the class of coherent GW signals of constant amplitude and polarization, which
include ``continuous GWs'', e.g.\ from non-axisymmetric spinning neutron stars (see
\cite{prix06:_cw_review} for a review), stellar-mass binary systems in the LISA frequency band
(e.g.\ \cite{krolak04:_optim_lisa,WhelanPrix08:_MLDC1B}), and coalescence of (non-precessing) binary
systems \cite{2001PhRvD..64d2004P}.

We distinguish two classes of signal parameters: (i) \emph{four} ``amplitude parameters'',
namely the amplitudes $\Ap$ and $\Ac$ of the two GW polarizations, the orientation angle $\psi$ of the
principal polarization axis, and the initial GW phase $\phi_0$, and (ii) the set of ``Doppler parameters''
$\Dop$, which determine the time evolution of the GW phase $\phi(t;\Dop)$. We restrict
our attention to \emph{targeted searches}, in which the Doppler parameters $\Dop$ are assumed to be
known, resulting in a detection problem with four unknown amplitude parameters.

The popular, yet \textit{ad-hoc}, orthodox approach consists of \emph{maximizing} the likelihood
function of the data over these four amplitude parameters.
It was first shown in \cite{jks98:_data} that this maximization can be achieved
analytically, resulting in a computationally very efficient detection statistic, known as the
$\F$-statistic, which has been used in a number of searches for GWs
(e.g. \cite{lsc04:_psr_j1939,lsc06:_coher_scorp_x,2005CQGra..22S1243A,PrixWhelan07:_MLDC1}).

Here we investigate an alternative Bayesian approach, which leads us to the Bayes factor as a
useful classical detection statistic (see also
\cite{2008PhRvD..78b2001V,2007PhRvD..76d3003C,2008arXiv0804.1161S,2008CQGra..25r4010V}).
Contrary to the maximum-likelihood approach, the Bayesian framework requires the explicit
prescription of a prior probability distribution for the unknown signal parameters.
We show that a particularly simple, yet unphysical, choice of amplitude prior results in the
$\F$-statistic as a special case of a Bayes factor.
This illustrates that frequentist \textit{ad-hoc} statistics often carry their own \emph{unchecked}
and \emph{implicit} priors, hidden from view and often unknown to the user (see also
\cite{2008CQGra..25k4038S,2008arXiv0809.2809S}).

We can derive a more natural amplitude prior from our model assumption about the emission of GWs from
non-axisymmetric spinning systems: the amplitude parameters are closely related
to the \emph{orientation} of the emitter with respect to the observer. In the absence of
astrophysical information, an isotropic probability distribution for the spin-axis orientation is
therefore the natural choice. We refer to the Bayes factor resulting from this amplitude prior as the ``$\B$-statistic''.
Isotropic spin-axis orientation priors have been used previously for Bayesian parameter estimation
\cite{lsc04:_psr_j1939,2005PhRvD..72j2002D,2004CQGra..21S1655U}, and in Monte-Carlo
simulations to determine frequentist upper limits \cite{lsc06:_coher_scorp_x,2007arXiv0708.3818L}.
A comparison of Bayesian and frequentist methods for setting upper limits can be found in
\cite{iraj_thesis_2007,lsc04:_psr_j1939}.

The $\F$-statistic (or matched filtering in general) has often been incorrectly referred to as
an ``optimal statistic'' (e.g.\
\cite{jks98:_data,lsc04:_psr_j1939,2001PhRvD..64d2004P,krishnan04:_hough,lsc06:_coher_scorp_x,PrixWhelan07:_MLDC1}
to name only a few).
Using Monte-Carlo simulations we show that the Bayesian $\B$-statistic is more powerful (i.e.\ has a
higher expected detection probability at equal false-alarm probability) than the
$\F$-statistic for GWs emitted by systems with random (isotropic)
spin-axis orientations. This is a direct consequence of the \mbox{$\B$-statistic} prior being consistent
with the injected distribution of parameters, contrary to the (implicit) $\F$-statistic
prior. Similar results were found previously in the case of burst detection statistics
\cite{2008arXiv0804.1161S,2008CQGra..25k4038S,2008arXiv0809.2809S}.

\section{Signal model: coherent gravitational waves of constant amplitudes}
\label{sec:param-family-grav}

The spatial metric perturbation $\htens(t)$ of a coherent GW of constant amplitudes (far from the
source) can be written as
\begin{equation}
  \label{eq:13}
  \htens(t) = \etens_+ \, \Ap\, \cos \left[ \phi(t;\Dop) + \phi_0\right]
  + \etens_\times \, \Ac \, \sin\left[\phi(t;\Dop) + \phi_0\right]\,,
\end{equation}
where  $\etens_+  = \uhat \otimes \uhat  - \vhat \otimes \vhat$, and
$\etens_\times = \uhat \otimes \vhat + \vhat \otimes \uhat$ are two polarization
basis tensors, constructed from a right-handed basis $\left\{\uhat ,\, \vhat,\, -\nhat \right\}$.
The unit-vector $\nhat$ is pointing along the line of sight from the detector to the source, and
the wave-plane basis vectors $\{\uhat,\,\vhat\}$ are aligned with the principal polarization axes of
the GW.
In general the GW phase $\phi(t;\Dop)$ depends on a set of \emph{Doppler parameters} $\Dop$, which
include the source sky-position $\nhat$, the GW frequency $\Freq$, and possibly higher-order time
derivatives of the frequency $\{\fdot,\,\fddot,\ldots\}$. If the source is a neutron star in a binary system, then $\Dop$
would also include the orbital parameters of the system.

We assume the GW emitter consists of a (non-axisymmetric) rotating system with spin $\vOm$ and
ellipticity $\eps$ with respect to the rotation axis.
The corresponding characteristic amplitude $h_0$ of the GW at the detector can be expressed as
\begin{equation}
  \label{eq:10}
  h_0 = \frac {4 G} {c^4} \frac {\Izz \,\Om^2}{d} \eps\,,
\end{equation}
where $\Izz$ is the moment of inertia with respect to the rotation axis, the rotation
rate is $\Om \equiv |\vOm|$, and $d$ is the distance to the detector.
\begin{figure}[htbp]
  \centering
  \includegraphics[width=0.5\textwidth]{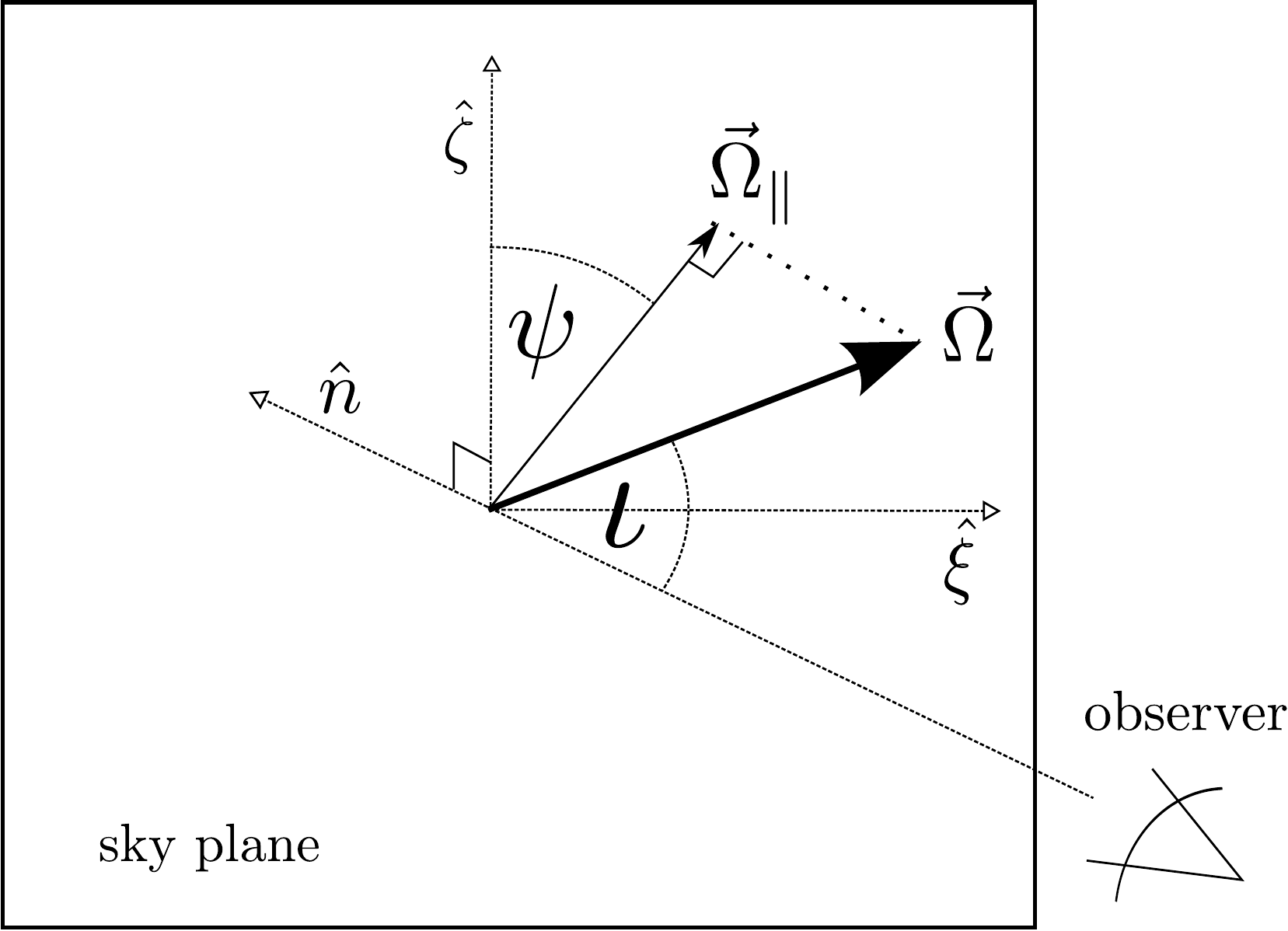}
  \caption{Source geometry angles: $\iota$ is the inclination angle between the
    rotation axis $\vOm$ and the line of sight ($-\nhat$). The polarization angle
    $\psi$ measures the orientation of the projected rotation axis $\vOm_{\parallel}$ in the sky
    plane with respect to an observer frame $\{\xihat,\zihat\}$.}
  \label{fig:angles}
\end{figure}
The emitter \emph{geometry} is fully characterized by two Euler angles describing the orientation of
the rotation axis, namely the \emph{inclination angle} $\iota$ of the rotation axis with respect to the
line of sight $\nhat$, and the \emph{polarization angle} $\psi$ describing the orientation of the projected
rotation axis in the sky-plane (see Fig.~\ref{fig:angles}).
%
The two polarization amplitudes $\Ap,\,\Ac$ in (\ref{eq:13}) can be expressed as
\begin{equation}
  \label{eq:12}
  \Ap = \frac{1}{2}h_0 ( 1 + \cosSQi )\,,\quad\mbox{and}\quad
  \Ac = h_0\,\cosi\,,
\end{equation}
in terms of the characteristic amplitude $h_0$ and the inclination angle $\iota$.
This corresponds to a choice of the wave-plane coordinate axis $\uhat$ perpendicular to the rotation
axis~\cite{bonazzola96:_gravit} 
and assigns $\Ap$ to the larger of the two principal polarization amplitudes, i.e.\ $\Ap \ge \Ac$.
%
%
For each sky position $\nhat$ we can define a source-independent (right-handed, orthonormal) basis
$\{\xihat,\,\zihat, -\nhat\}$, e.g.\ where $\xihat$ lies in the ecliptic plane and $\zihat$ in the
northern hemisphere.
This allows us to define the polarization angle $\psi$ as the angle between the principal
polarization axis $\uhat$ of the GW and the basis vector $\xihat$,
i.e.\ $\psi \equiv \measuredangle(\uhat,\xihat) = \measuredangle(\vhat,\zihat)$, where
$\vhat \propto \vOm_{\parallel}$ (see Fig.~\ref{fig:angles}).
%
%
As first shown in \cite{jks98:_data}, the strain $h(t)$ measured in the detector due to a GW signal
(\ref{eq:13}) can be expressed as
\begin{equation}
  \label{eq:20}
  h(t;\Amp,\Dop) = \Amp^\mu\, h_\mu(t;\Dop)\,,
\end{equation}
where we use automatic summation $\sum_{\mu=1}^4$ over repeated amplitude indices $\mu$.
The explicit form of the four ``basis functions'' $h_\mu(t;\Dop)$ can be found in
\cite{PrixWhelan07:_MLDC1}, for example, but is not important for the following discussion.
The four canonical amplitudes $\Amp^\mu$ are defined as
\begin{equation}
  \eqalign{
    \Amp^1 &= \;\;\Ap\,\cos 2\psi\,\cos\phi_0 - \Ac\,\sin 2\psi\,\sin\phi_0\,,\\
    \Amp^2 &= \;\;\Ap\,\sin 2\psi\,\cos\phi_0 + \Ac\,\cos 2\psi\,\sin\phi_0\,,\\
    \Amp^3 &= -\Ap\,\cos 2\psi\,\sin\phi_0 - \Ac\,\sin 2\psi\,\cos\phi_0\,,\\
    \Amp^4 &= -\Ap\,\sin 2\psi\,\sin\phi_0 + \Ac\,\cos 2\psi\,\cos\phi_0\,.
  }\label{eq:14}
\end{equation}
The set of \emph{amplitude parameters} $\Amp$ can therefore be expressed either in ``physical
coordinates'' $\AmpPhys^\mubar$, i.e.\
\begin{equation}
  \label{eq:9}
  \{\Amp\}^\mubar = \AmpPhys^i = \{h_0, \cosi, \psi, \phi_0\}\,,
\end{equation}
or in ``canonical coordinates'' $\Amp^\mu$ given in (\ref{eq:14}).

\section{Hypothesis testing for GW detection}
\label{sec:GWstatistics}

\subsection{Simple versus composite hypotheses}
\label{sec:hypothesis-testing}

In the following we restrict ourselves to the case where we need to decide only between two hypotheses, namely
$\Hyp_\Noise \equiv$ ``the data $x$ consist of only noise $n$'', and
$\Hyp_\Signal \equiv$ ``the data contains a signal $s$ in addition to noise $n$'',
where we assume a signal $s = h(t; \Amp,\Dop)$ of the form (\ref{eq:20}).
For simplicity we assume the Doppler parameters $\Dop = \Dop_\sig$ to be known \textit{a-priori},
while the four amplitude parameters $\Amp \in \AmpSpace$ are unknown.
This corresponds to a \emph{targeted} search for GWs, for example from an isolated pulsar with known
sky position and GW frequency $\Freq(t)$.
We can formally write the two hypothesis as
\begin{equation}
  \eqalign{
    \Hyp_\Noise: x(t) &= n(t)\,,\\
    \Hyp_\Signal: x(t) &= n(t) + s(t; \Amp,\Dop_\sig)\,,\quad \mbox{for any}\quad\Amp \in \AmpSpace\,.
  } \label{eq:4}
\end{equation}
Note that $\Hyp_\Noise$ is a \emph{simple} hypothesis, which means that all model parameters are
fully specified, namely $s=0$. On the other hand, $\Hyp_\Signal$ is a \emph{composite} hypothesis,
as the amplitude parameters $\Amp\in \AmpSpace$ of the signal are unknown.
The composite hypothesis $\Hyp_\Signal$ can be considered as a union of simple hypotheses, i.e.\
$\Hyp_\Signal = \underset{\Amp\in\AmpSpace}\cup \Hyp_\Signal(\Amp)$.

\subsection{Scalar product and Gaussian noise}
\label{sec:scal-prod-gauss}

Assuming Gaussian stationary noise with  known (single-sided) power spectral density $\Sn$, the
probability density function (pdf) for a particular pure-noise time series $x = n$ can be written as
\begin{equation}
  \label{eq:24}
  \pdf\left(n | \Sn \right) = \kappa \, e^{-\frac{1}{2}\scalar{n}{n}}\,,
\end{equation}
where $\kappa$ is a normalization constant and we defined the scalar product $\scalar{x}{y}$
between time-series $x(t)$ and $y(t)$ as
\begin{equation}
  \label{eq:25}
  \scalar{x}{y} \equiv 4 \Re \int_0^\infty \frac{\tilde{x}(f)\, \tilde{y}^*(f)}{\Sn(f)}\,d f\,,
\end{equation}
where $\tilde{x}(f)$ denotes the Fourier transform of $x(t)$ and $^*$ denotes complex conjugation
(e.g.\ see \cite{finn92:_detection}).
The likelihood of observing data $x(t)$ in the presence of a signal $s(t)$ is therefore
\begin{equation}
  \label{eq:15}
  \pdf\left(x | s\,\Sn \right) = \kappa \, e^{-\frac{1}{2}\scalar{(x-s)}{(x-s)} }\,.
\end{equation}

\subsection{Neyman-Pearson optimality}
\label{sec:neym-pears-optim}

A \emph{detection statistic} $\Stat(x)$ is a real-valued function of the data $x$, such that the
corresponding \emph{test} of threshold $\thresh{\Stat}$ decides for $\Hyp_\Noise$ if
$\Stat(x) < \thresh{\Stat}$, and for $\Hyp_\Signal$ if $\Stat(x) > \thresh{\Stat}$.
Such a test is typically subject to two types of error: a ``false alarm'',
i.e.\ choosing $\Hyp_\Signal$ when $\Hyp_\Noise$ is true, and a ``false dismissal'',
i.e.\ rejecting $\Hyp_\Signal$ when it is in fact true.
We can express the probability $\fA$ of a false alarm as
\begin{equation}
  \label{eq:7}
  \fA(\thresh{\Stat}) = P(\Stat>\thresh{\Stat}|\Hyp_\Noise) = \int_{\thresh{\Stat}}^\infty \pdf(\Stat|\Hyp_\Noise ) \, d\Stat\,.
\end{equation}
The probability $\fD$ of a false dismissal is
$\fD(\thresh{\Stat};\pA) = P(\Stat < \thresh{\Stat}|\Hyp_{\Signal}\,\pA)$, where we notice
that the specification (\ref{eq:4}) of the signal hypothesis $\Hyp_\Signal$ is in fact \emph{incomplete}:
in addition to the condition $\Amp \in \AmpSpace$, we also need to specify the
probability density for $\Amp$, which we denote as $\pA$.
When measuring $\fD$ in a Monte-Carlo simulation, $\pA$ would corresponds to the assumed underlying
population, from which signals are randomly drawn in each test.
The complementary \emph{detection probability} $\eta \equiv 1 - \fD$ is
\begin{equation}
  \label{eq:8}
  \eta(\thresh{\Stat};\pA) = \int_{\thresh{\Stat}}^{\infty} \pdf(\Stat|\Hyp_{\Signal}\,\pA) \, d\Stat\,.
\end{equation}
Note that this contains the usual definition of the \emph{power function}
$\eta(\thresh{\Stat};\Amp)$ as a special case, where $\pA = \Amp$. The definition (\ref{eq:8}) can also
be interpreted as the \emph{expected} power over a population $\pA$.
The Neyman-Pearson framework for hypothesis testing defines the \emph{most powerful test} of size $\fA$
as a test that has the highest detection probability $\eta$ (i.e.\ smallest false dismissal $\fD$)
for a false-alarm probability of at most $\fA$ (e.g.\ see \cite{helstrom68:_signal_detection,1999kats.book.....S}).
In this framework one compares the respective detection probabilities $\eta$ of different
detection statistics at a given false-alarm probability $\fA$, which defines the
\emph{receiver-operator characteristics} (ROC), namely the function $\eta(\fA;\pA)$.

In general the relative performance of different detection statistics will depend on the assumed
probability distribution $\pA$ of signal parameters. One statistic can be more efficient in certain
parts of parameter space and less efficient in others. If a test is most powerful over the whole
parameter space (i.e.\ has the highest $\eta(\thresh{\Stat};\Amp)$ for all $\Amp \in \AmpSpace$), it
a called a \emph{uniformly most powerful} test.

\section{Frequentist maximum-likelihood approach: the $\F$-statistic}
\label{sec:maxim-likel-detect}

When comparing two \emph{simple} hypotheses, such as $\Hyp_\Noise$ and $\Hyp_\Signal(\Amp)$
\emph{for known} $\Amp$, the Neyman-Pearson Lemma states that the most powerful test (cf.\
section~\ref{sec:neym-pears-optim}) is the \emph{likelihood-ratio} $\Lr$, defined as
\begin{equation}
  \label{eq:5}
  \Lr(x;\,\Amp) \equiv \frac{\pdf(x|\Hyp_\Signal(\Amp))}{\pdf(x|\Hyp_\Noise)}\,.
\end{equation}
Assuming Gaussian noise and using (\ref{eq:15}), we explicitly obtain
\begin{equation}
  \label{eq:6}
  \Lr(x;\,\Amp) = \exp\left[\scalar{x}{s} - \frac{1}{2}\scalar{s}{s}\right]\,.
\end{equation}
However, in the case of a \emph{composite} hypothesis $\Hyp_\Signal$ with \emph{unknown} amplitude
parameters $\Amp$, the orthodox frequentist framework does not generally provide a canonical
detection statistic. Interestingly, one cannot even define a frequentist likelihood $\pdf(x|\Hyp_\Signal)$.
%
A common but \textit{ad-hoc} approach to dealing with composite hypotheses consists of using
the \emph{maximum} of the likelihood ratio $\Lr(x;\Amp)$ over the parameter
space $\AmpSpace$, i.e.\ define
\begin{equation}
  \label{eq:31}
  \Lr_\ML(x) \equiv \underset{\Amp \in \AmpSpace}{\max} \, \Lr(x;\,\Amp)
\end{equation}
as a composite-hypothesis test: decide for $\Hyp_\Signal$ if $\Lr_\ML(x) > \thresh{\Lr}$ and
$\Hyp_\Noise$ otherwise.
%
Using (\ref{eq:6}) and (\ref{eq:20}), the likelihood-ratio function can be written more explicitly as
\begin{equation}
  \label{eq:32}
  \Lr(x;\,\Amp) = \exp\left[ \Amp^\mu \,x_\mu - \frac{1}{2} \Amp^\mu \M_{\mu\nu} \Amp^\nu \right]\,,
\end{equation}
where we defined
\begin{equation}
  \label{eq:33}
  x_\mu \equiv \scalar{x}{h_\mu(\Dop_\sig)}\,,\quad
  \M_{\mu\nu} \equiv \scalar{h_\mu(\Dop_\sig)}{h_\nu(\Dop_\sig)}\,.
\end{equation}
We see that $\Lr(x;\Amp)$ is a Gaussian function in $\Amp^\mu$, so we can analytically maximize it
to obtain
\begin{eqnarray}
  \label{eq:34}
  \Lr_\ML(x) &= e^{\F(x)}\,,\quad\mbox{with}\quad
  \F(x) \equiv \frac{1}{2} \,x_\mu \,\M^{\mu\nu} \,x_\nu\,,
\end{eqnarray}
with the inverse matrix $\M^{\mu\nu}$ defined via $\M^{\mu\sigma}\M_{\sigma\nu} = \delta^\mu_\nu$.
This defines the so-called $\F$-statistic, which was first derived in this context in \cite{jks98:_data}.
The statistic $2\F(x)$ can be shown to be $\chi^2$-distributed with four degrees of freedom, and a
non-centrality parameter $\rho^2 \equiv \scalar{s}{s}$, where $\rho$ is called the
(optimal) \emph{signal-to-noise ratio}. The expectation value of $2\F$ is $E[2\F] = 4 + \rho^2$.

\section{Bayesian hypothesis testing}

The Bayesian hypothesis-testing framework follows uniquely from a straightforward application of
the probability axioms (cf \cite{jaynes:_logic_of_science,sivia96:_bayesian,2008PhRvD..78b2001V,2005blda.book.....G}).
For any question of interest one can (at least in principle) \emph{compute} the probability of
different hypotheses, optimally using the available information such as the observed data $x(t)$,
and all our prior information and assumptions, which we denote by '$I$'.
Here we use the Bayesian approach to construct a classical \emph{detection statistic}, in
order to compare its performance to the frequentist $\F$-statistic in the Neyman-Pearson framework.

For any hypothesis $\Hyp_i$ we can directly express the probability of $\Hyp_i$ being true given
the data $x$ and our background assumptions $I$, namely
\begin{equation}
  \label{eq:36}
  P(\Hyp_i | x \, I) = \frac{\pdf(x | \Hyp_i\, I ) \, P(\Hyp_i | I )}{\pdf(x | I )}\,.
\end{equation}
This expression is known as \emph{Bayes' theorem}, and it follows directly from the product rule of
probabilities applied to $\pdf(\Hyp_i \, x | I )$.
The term $P(\Hyp_i|I)$ is the \emph{prior probability} for $\Hyp_i$. Contrary to the orthodox
frequentist framework, the (marginal) likelihood\footnote{also known as the \emph{evidence}} $\pdf(x |
\Hyp_i \,I)$ of observing data $x$ given $\Hyp_i$  is well-defined even for composite hypotheses.
In order to compute $\pdf(x|\Hyp_\Signal\, I)$, we simply use the product rule to write
\begin{equation}
  \label{eq:1}
  \pdf(\Amp | x \,\Hyp_\Signal\,I) = \frac{\pdf(\Amp \, x | \Hyp_\Signal\,I)}{\pdf(x|\Hyp_\Signal\,I)}\,,
\end{equation}
and invoking the normalization condition $\int \pdf(\Amp|x\Hyp_\Signal\,I)\,d^4\Amp = 1$, we obtain
\begin{equation}
  \eqalign{
    \pdf(x | \Hyp_\Signal \, I) &= \int_{\AmpSpace} \pdf(\Amp\,x| \Hyp_\Signal\,I)\, d^4\Amp \\
    &= \int_{\AmpSpace} \pdf( x | \Amp\, \Hyp_\Signal \, I)\, \pdf(\Amp | \Hyp_\Signal \, I)\, d^4\Amp\,.
  }   \label{eq:37}
\end{equation}
%
If $\{\Hyp_i\}_{i=1}^m$ is a set of $m$ mutually exclusive and exhaustive hypotheses, i.e.\ exactly
one of them is true, then one obtains the normalization condition $\sum_{i=1}^m P(\Hyp_i|x I)=1$,
which determines the denominator $\pdf(x|I)$ in (\ref{eq:36}).
We do not need to make this assumption, however, as we can instead compute the \emph{relative}
probability of $\Hyp_\Signal$ with respect to $\Hyp_\Noise$, which is known as the
(posterior) \emph{odds ratio} $O_{\Signal\Noise}$, namely
\begin{equation}
  \label{eq:40}
  O_{\Signal\Noise}(x|I) \equiv \frac{P(\Hyp_\Signal|x\,I)}{P(\Hyp_\Noise|x\,I)}
  = \frac{\pdf(x|\Hyp_\Signal I)}{\pdf(x|\Hyp_\Noise I)}\,\frac{P(\Hyp_\Signal|I)}{P(\Hyp_\Noise|I)}\,.
\end{equation}
This expressions shows how the prior odds ratio $P(\Hyp_\Signal|I)/P(\Hyp_\Noise|I)$ gets
``updated'' by the observation of $x$, namely by multiplication with the (marginal) \emph{likelihood-ratio}
\begin{equation}
  \label{eq:41}
  B_{\Signal\Noise}(x|I) \equiv \frac{\pdf(x|\Hyp_\Signal\,I)}{\pdf(x|\Hyp_\Noise\,I)}\,,
\end{equation}
which is also known as the \emph{Bayes factor}.
Note that the prior odds ratio is a constant factor in $O_{\Signal\Noise}$, and therefore plays no role
in constructing a classical \emph{detection statistic} $\Stat(x)$ (any monotonic function of
$\Stat(x)$ has the same power).
Using (\ref{eq:37}) and (\ref{eq:5}), we can write the Bayes factor (\ref{eq:41}) explicitly as
\begin{equation}
  \label{eq:2}
    B_{\Signal\Noise}(x|I) 
    = \int_\AmpSpace \Lr(x;\Amp)\, \pdf(\Amp|\Hyp_\Signal\,I)\,d^4\Amp\,.
\end{equation}
Note that while the $\F$-statistic (\ref{eq:34}) was obtained by \emph{maximizing} the likelihood
ratio $\Lr(x;\Amp)$ over the ``nuisance parameters'' $\Amp^\mu$, the Bayes factor
$B_{\Signal\Noise}(x)$ consists of \emph{marginalizing} $\Lr(x;\Amp)$ with an amplitude prior
$\pdf(\Amp|\Hyp_\Signal\,I)$. In order to uniquely specify the Bayes factor, we therefore need to
determine the function $\pdf(\Amp|\Hyp_\Signal I)$, which adequately describes our ignorance of the
pulsar amplitude parameters $\Amp$.

\subsection{Uniform priors in $\Amp^\mu$-coordinates: rediscovering the $\F$-statistic}
\label{sec:using-flat-priors}

Considering the form (\ref{eq:32}) of $\Lr(x;\Amp)$, a straightforward choice would be uniform priors in
coordinates $\Amp^\mu$. We refer to this as the ``canonical prior'' $\Prior_\canonical$, namely
\begin{equation}
  \label{eq:47}
  \pdf(\{\Amp^\mu\}| \Hyp_\Signal\,\Prior_\canonical\, I) = \left\{ \begin{array}{l l}
      C & \quad \mbox{if } h_0(\{\Amp^\mu\}) < h_0^{\max}\,,\\
      0 & \quad \mbox{otherwise}\,,
    \end{array}\right.
\end{equation}
where $h_0^{\max}$ is the maximum amplitude we consider possible, and $h_0(\{\Amp^\mu\})$ is given
by inversion of Eq.~(\ref{eq:14}). The normalization constant $C$ is determined by
$\int_{\AmpSpace}\pdf(\Amp|\Hyp_\Signal\,I)\,d^4\Amp = 1$. The actual choice of $h_0^{\max}$
is unimportant for the properties of $B_{\Signal\Noise}(x)$ \emph{as a detection statistic}, because
for large $h_0^{\max}\gg 1$, the marginalization (\ref{eq:2}) leads to a Gaussian integral, namely
\begin{equation}
  \label{eq:38}
  \eqalign{
    B_{\Signal\Noise}(x|\Prior_\canonical\,I) &= C \int_{\AmpSpace} \Lr(x;\Amp)\,d^4\Amp\\
    &= C\, (2\pi)^2 (\det\M)^{-1/2}\, e^{\F(x)}\,,
  }
\end{equation}
where $\det\M$ is the determinant of the matrix $\M_{\mu\nu}$.
We see that uniform amplitude priors in $\Amp^\mu$-coordinates lead us back to the $\F$-statistic
(\ref{eq:34}). However, there is an additional antenna-pattern factor
$(\det\M)^{-1/2}$, which depends on the sky-position $\nhat$ and the observation period.
For a targeted search with a single known sky position, this is a constant
factor which does not affect the power of the detection statistic, i.e.\
$B_{\Signal\Noise}(x|\Prior_\canonical\,I)$ is equivalent to $\F(x)$.
This weighting factor would play a role, however, when investigating searches over unknown sky
position.
A similar effect was first noted in the Bayesian analysis of burst detection statistics \cite{2008arXiv0809.2809S}.

\subsection{Physical priors in amplitude-space: introducing the $\B$-statistic}
\label{sec:what-are-physical}

Despite our assumed ``ignorance'' about the amplitude parameters of the GW signal, we have made a
number of model assumptions about the geometry of the emitting system (see section~\ref{sec:param-family-grav}).
In a sense the physical model describing the emitting system singles out a
preferred coordinate system in $\AmpSpace$, in which we should express our ignorance. We refer
to the resulting prior as the ``physical prior'' $\Prior_\physical$.

The initial phase $\phi_0$ is directly related to the rotation angle of the quadrupolar deformation
with respect to the rotation axis $\vOm$ at some fixed reference time. The probability distribution
for $\phi_0$ is therefore independent of $\{h_0,\cosi,\psi\}$, and by rotational symmetry we can
assign a uniform prior, i.e.\
\begin{equation}
  \label{eq:48}
  \pdf(\phi_0| \Prior_\physical \, I) = \frac{1}{2\pi}\,,\quad \phi_0 \in [0, 2\pi)\,.
\end{equation}
In section~\ref{sec:param-family-grav} and Fig.~\ref{fig:angles} we have seen that $\cosi$ and $\psi$
determine the orientation of the rotation axis $\vOm$ with respect to the observer frame
$\{\xihat,\zihat,-\nhat\}$.
If we have no information about the orientation of the emitting system, then rotational symmetry
dictates an isotropic probability distribution for $\vOm$.
The surface element on the unit sphere of $\vOm$ orientations can be expressed as
$d^2S = |d\cosi|\,|d\psi|$, and because $\cosi$ and $\psi$ are independent degrees of freedom, their
respective prior probabilities are
\begin{eqnarray}
  \label{eq:51}
  \pdf(\cosi\,|\Prior_\physical \,I) &= \frac{1}{2}\,,\qquad \cosi\in[-1,1]\,,\\
  \pdf(\psi\,|\Prior_\physical \, I)  &= \frac{2}{\pi}\,,\qquad \psi \in [-\pi/4, \pi/4)\,,
\end{eqnarray}
where we used the fact that gauge transformations
$\{\psi\rightarrow \psi + \pi/2,\, \phi_0\rightarrow\phi_0+\pi\}$ leave the observed signal
(\ref{eq:20}) unchanged, so $\psi$ can always be brought into the range $\psi \in [-\pi/4,\,\pi/4)$.
Note that these priors are identical to those used previously in Bayesian parameter estimation
\cite{lsc04:_psr_j1939,2005PhRvD..72j2002D,2004CQGra..21S1655U} and Monte-Carlo
simulations for frequentist upper limits  \cite{lsc06:_coher_scorp_x,2007arXiv0708.3818L}.

Contrary to the angle variables $\cosi,\psi$ and $\phi_0$, there is no unique natural choice of
uninformed prior for the amplitude $h_0$. One could derive a prior for $h_0$ from
Eq.~(\ref{eq:10}), if astrophysical priors for the deformation $\eps$, spin rate $\Om$ and
distance $d$ are available. Other possibilities include a ``maximum entropy'' prior, or a Jeffreys
prior.
For simplicity, however, we simply chose a uniform prior, namely
\begin{equation}
  \label{eq:52}
  \pdf(h_0|\Prior_\physical\,I) = \frac{1}{h_0^{\max}}\,,\qquad h_0 \in [ 0, h_0^{\max}]\,.
\end{equation}
Combining (\ref{eq:48}) - (\ref{eq:52}) we obtain an amplitude prior of the form
\begin{equation}
  \label{eq:54}
  \pdf(h_0,\cosi,\psi,\phi_0\,|\Prior_\physical\,I) = \frac{1}{2 \pi^2\,\,h_0^{\max}} = C'\,,\quad h_0\in[0,h_0^{\max}]\,,
\end{equation}
which for simplicity of notation we refer to as the physical prior, while this
qualifier can only be justified for the angle variables.
%
Substituting the prior $\Prior_\physical$ in the Bayes factor (\ref{eq:2}), and assuming
$h_0^{\max}\gg 1$, we now obtain
\begin{equation}
  \label{eq:3}
  B_{\Signal\Noise}(x|\Prior_\physical\,I) = C'\, \int_0^{\infty} \!\!dh_0 \int_{-1}^{1}\!\!d\!\cosi
  \int_{-\frac{\pi}{4}}^{\frac{\pi}{4}} \!\!d\psi \int_{0}^{2\pi} \!\! d\phi_0 \,\, \Lr(x;\,\Amp)\,,
\end{equation}
with the likelihood-ratio $\Lr(x;\Amp)$ of Eq.~(\ref{eq:32}). We use this Bayes factor as a new
classical \emph{detection statistic} $\B(x)$, which we refer to as the ``$\B$-statistic'',
namely
\begin{equation}
  \label{eq:28}
  \B(x) \equiv B_{\Signal\Noise}(x|\Prior_\physical\,I)\,.
\end{equation}

\subsection{Relation between amplitude priors $\Prior_\canonical$ and $\Prior_\physical$}
\label{sec:relat-betw-canon}

In order to compare the physical amplitude prior $\Prior_\physical$ of
Eq.~(\ref{eq:54}) to the canonical prior $\Prior_\canonical$ of Eq.~(\ref{eq:47}), we use the
coordinate transformation (\ref{eq:14}) relating $\Amp^\mu$ and $\AmpPhys^\mubar$.
The Jacobian $J$ of this transformation is found as
\begin{equation}
  \label{eq:53}
  J \equiv \left|\det\left( \frac{\partial \Amp^\mu}{\partial \AmpPhys^\mubar} \right)\right|
  = \frac{h_0^3}{4} \, \left( 1 - \cosSQi \right)^3\,.
\end{equation}
Using the identity
\begin{equation}
  \label{eq:57}
  \pdf\left(\{\Amp^\mu\}\,|\,\Prior_\canonical\,I\right) \, d^4\!\Amp = \pdf\left(\{\AmpPhys^\mubar\}\,|\,\Prior_\canonical\,I\right)\, d^4\!\AmpPhys\,,
\end{equation}
together with the relation $d^4\!\Amp = J\,d^4\!\AmpPhys$ between volume elements, we can
translate $\Prior_\canonical$ into physical coordinates $\{\AmpPhys^\mubar\} = \{h_0,\cosi,\psi,\phi_0\}$,
namely
\begin{equation}
  \label{eq:56}
  \pdf\left(h_0,\cosi,\psi,\phi_0\,|\,\Prior_\canonical\,I\right) = \frac{C}{4}\,h_0^3\,\left(1 - \cosSQi\right)^3\,,
\end{equation}
which can be compared to the physical prior $\Prior_\physical$ in Eq.~(\ref{eq:54}).
We see that $\Prior_\canonical$ agrees with $\Prior_\physical$ in assigning uniform prior
probabilities to $\phi_0$ and $\psi$, but the prior densities on $\cosi$ and $h_0$ are
very different.

The canonical prior information $\Prior_\canonical$, which is \emph{implicit} in
the $\F$-statistic (cf.\ section~\ref{sec:using-flat-priors}), is therefore found to be rather
unphysical: a higher prior probability is assigned to stronger signals compared to weaker ones, and
signals with near-linear polarization ($\cosi \sim 0$, corresponding to ``edge-on'' emitters) are
given undue weight compared to signals with near-circular polarization ($|\cosi|\sim 1$,
corresponding to ``face-on'' emitters). This amounts to postulating a non-isotropic
probability distribution for the orientation $\vOm$ of spinning GW sources, such that $\vOm$ favors
orientations orthogonal to the line of sight $\nhat$.

\section{Comparing detection efficiencies of $\F(x)$ and $\B(x)$}
\label{sec:comp-detect-effic}

\subsection{Estimating the ROC curves}
\label{sec:estim-roc-curv}

We use the classical Neyman-Pearson framework (cf.\ section~\ref{sec:neym-pears-optim}) in order to
compare the detection efficiency, or ``power'', of the $\F$-statistic (\ref{eq:34}) and the
$\B$-statistic (\ref{eq:3}).
The practical Monte-Carlo procedure for estimating the ROC curve $\eta(\fA;\pA)$ for any detection
statistic $\Stat$ is straightforward: first generate a large sample of $\NMC$ random draws
$\{\Stat_\Noise\}$ of the statistic $\Stat$ for the case of \emph{no signal}, i.e.\ $s=0$.
From this distribution we can estimate the false-alarm probability (\ref{eq:7}) as a function of
the threshold $\thresh{\Stat}$, namely
\begin{equation}
  \label{eq:83}
  \fA(\thresh{\Stat}) \approx \frac{N_{\Stat_\Noise > \thresh{\Stat}}}{\NMC}\,,
\end{equation}
where $N_{\Stat > \thresh{\Stat}}$ is the number of $\Stat_\Noise$ values found above the threshold
$\thresh{\Stat}$.
Similarly, in the \emph{signal case} $\Hyp_\Signal$, we randomly draw signal parameters from the
assumed population $\pA$ and generate corresponding random draws $\{\Stat_{\Signal}\}$ of the
statistic $\Stat$. From this distribution we can estimate the detection probability (\ref{eq:8}), namely
\begin{equation}
  \label{eq:84}
  \eta(\thresh{\Stat};\pA) \approx \frac{N_{\Stat_{\Signal} > \thresh{\Stat}}}{\NMC}\,.
\end{equation}
Inverting (\ref{eq:83}) to yield $\thresh{\Stat}(\fA)$ and substituting this into (\ref{eq:84}),
we obtain the ROC curve $\eta(\fA;\pA)$.

\subsection{Parameters used in Monte-Carlo simulation}
\label{sec:param-used-monte}

In targeted searches the Doppler parameters $\Dop$ of the signal are known, and for simplicity of
this example we fixed these parameters as: right ascension $\alpha = 2\,\rad$, declination
$\delta=-0.5\,\rad$ and a constant frequency without spindown. We assume the detector location to be
LIGO Hanford, and an observation with GPS start time of $t_0=756950413$ and duration of $T=25\,$hours.
The resulting numerical components of the antenna-pattern matrix $\M_{\mu\nu}$ of Eq.~(\ref{eq:33})
are found as: $\M_{11} = \M_{33} = \frac{T}{\Sn}\,A$, $\M_{22}=\M_{44}=\frac{T}{\Sn}\,B$, and
$\M_{12} = \M_{34}=\frac{T}{\Sn}\,C$, with $A = 0.154$, $B = 0.234$ $C = -0.0104$, with all other
components (approximately) zero.
These parameters are given for the sake of completeness, the qualitative conclusions do not depend
on these choices.
We used $\NMC = 10^6$ random draws for each distribution, and we estimate the errors on
$\eta(\fA;\pA)$ using a jackknife estimator (see \cite{conway:Voronoi1984}) with $100$
subsets. The estimated $1\sigma$ errors on the detection probability in the following ROC curves are
always less than $\sigma(\eta) < 0.004$.

\subsection{Monte-Carlo results}

Because $\Hyp_\Signal$ is a composite hypothesis, the ROC curves depend on the choice of
injected signal population $\pA$. In order to illustrate the dependency on the amplitude
parameter space, we first consider two highly unphysical choices of signal populations, namely
(i) $\pA$ consisting of a \emph{single}, linearly polarized signal with $\cosi=0$, $\psi=0$
and
(ii) a \emph{single}, (nearly) circularly polarized signal with $\cosi=0.99$, $\psi=0$.
In both cases we fixed the SNR of the signal to be $\rho=4$.
Note that the choice of $\phi_0$ is irrelevant for both $\F$ and $\B$.
These two choices reflect universes in which all spinning GW sources happen to be (i) edge-on or
(ii) face-on, without the observer having any knowledge about it.
\label{sec:monte-carlo-results}
\begin{figure}[h!tbp]
  \mbox{
    \hspace*{-0.5cm}
    \includegraphics[width=0.55\columnwidth,clip]{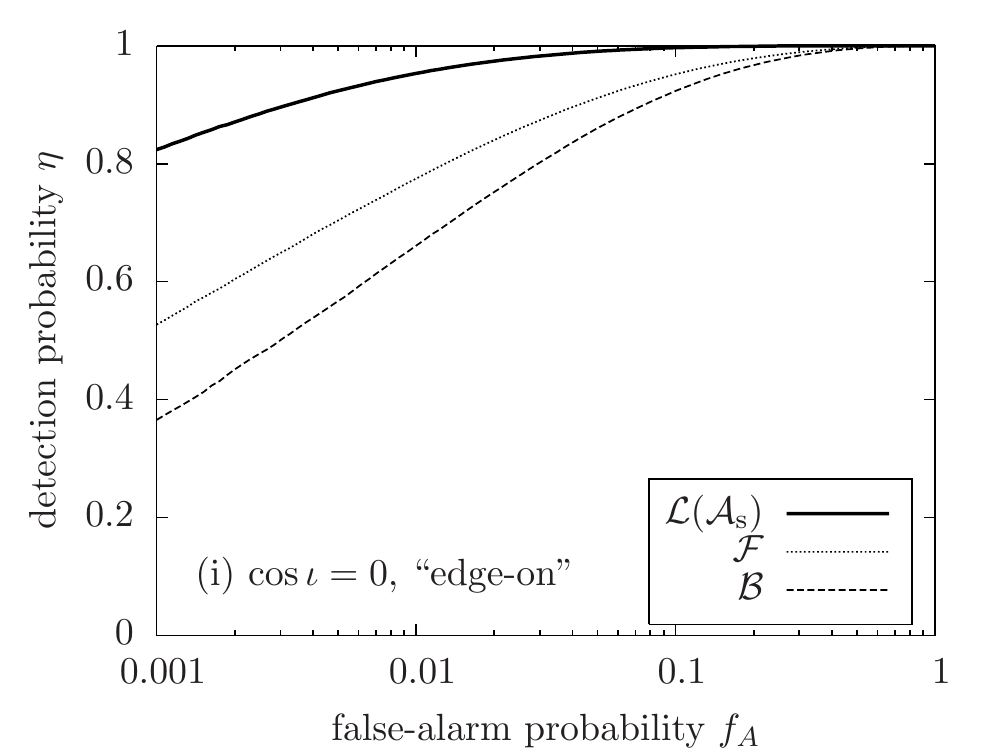}
    \hspace*{-0.8cm}
    \includegraphics[width=0.55\columnwidth,clip]{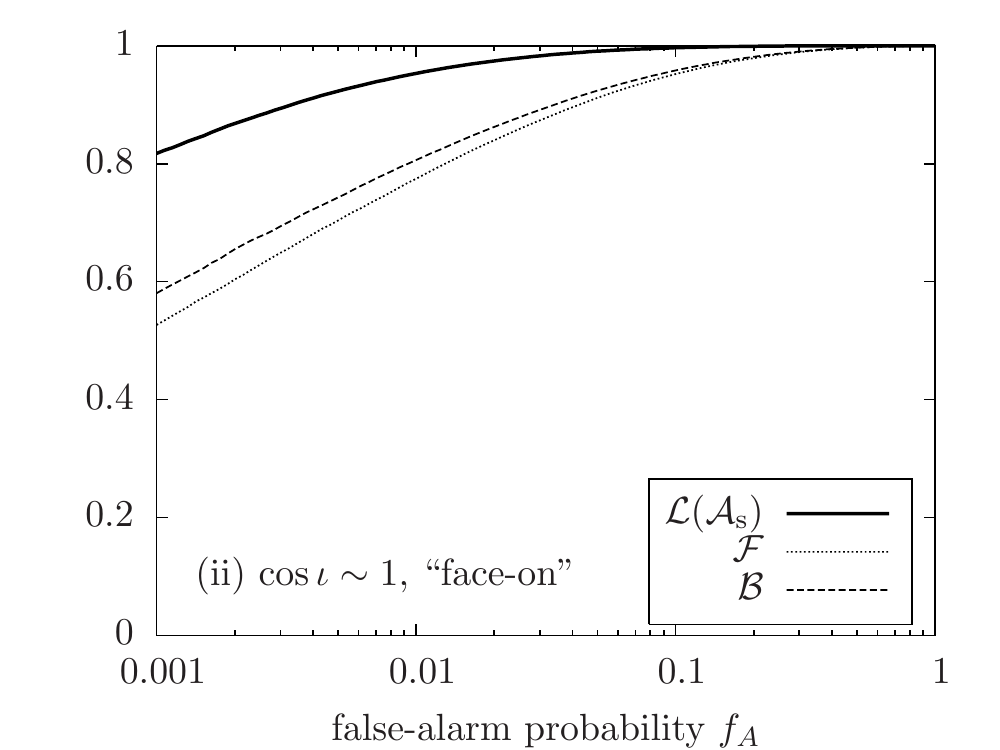}
  }
  \caption{ROC curves $\eta(\fA;\pA)$ comparing $\F$-statistic (\ref{eq:34}), $\B$-statistic
    (\ref{eq:28}), and the perfect-match likelihood ratio $\Lr(\Amp_\sig)$.
    The chosen signal populations $\pA$ have fixed SNR of $\rho=4$, and consist of
    (i) a \emph{single} linearly polarized signal with $\cosi=0$, $\psi=0$ (\textit{left panel}), and
    (ii) a (nearly) circularly polarized signal with $\cosi=0.99$, $\psi=0$ (\textit{right panel}).
  }
  \label{fig:roc_example1}
\end{figure}
The results of these simulations are shown in Fig.~\ref{fig:roc_example1}. For comparison purposes
we also plot the perfect-match likelihood ratio $\Lr(\Amp_\sig)$, which is optimal by the
Neyman-Pearson lemma for testing the simple hypothesis $\Hyp_\Signal(\Amp_\sig)$  (cf.\ section
\ref{sec:neym-pears-optim}), but requires all signal parameters $\Amp_\sig$ to be exactly known
beforehand.

We see that the $\F$-statistic is more powerful than the $\B$-statistic if the signal is
linearly-polarized, while the $\B$-statistic dominates for (near-) circularly-polarized GWs.
This is not surprising given that the implicit $\F$-statistic prior $\Prior_\canonical$ is biased in
favor of linear polarization (cf.\ section~\ref{sec:relat-betw-canon}).
The results in Fig.~\ref{fig:roc_example1} show that neither $\F$- nor $\B$-statistic is
\emph{uniformly most powerful} (cf.\ section~\ref{sec:neym-pears-optim}) over the amplitude
parameter space $\AmpSpace$.
Note that this does \emph{not} imply that the $\F$-statistic is more powerful if we \emph{know}
a given source to be (near-) linearly polarized. One would fold this knowledge into the prior
$\Prior$ in the Bayes factor (\ref{eq:2}), while there is no natural way in which this knowledge can
be incorporated into the $\F$-statistic. The resulting Bayes factor would therefore be more powerful
than the $\F$-statistic.

In the next step we look at more realistic situations in which the injected signals are drawn from
a population $\pA$ of randomly distributed $\cosi$ and $\psi$, according to the physical prior
$\Prior_{\physical}$ (cf.\ section~\ref{sec:what-are-physical}), with (iii) a \emph{fixed SNR} of
$\rho=4$ and (iv) a \emph{fixed amplitude} of $h_0=10\,\sqrt{\Sn}$.
\begin{figure}[htbp]
  \mbox{
    \hspace*{-0.5cm}
    \includegraphics[width=0.55\columnwidth,clip]{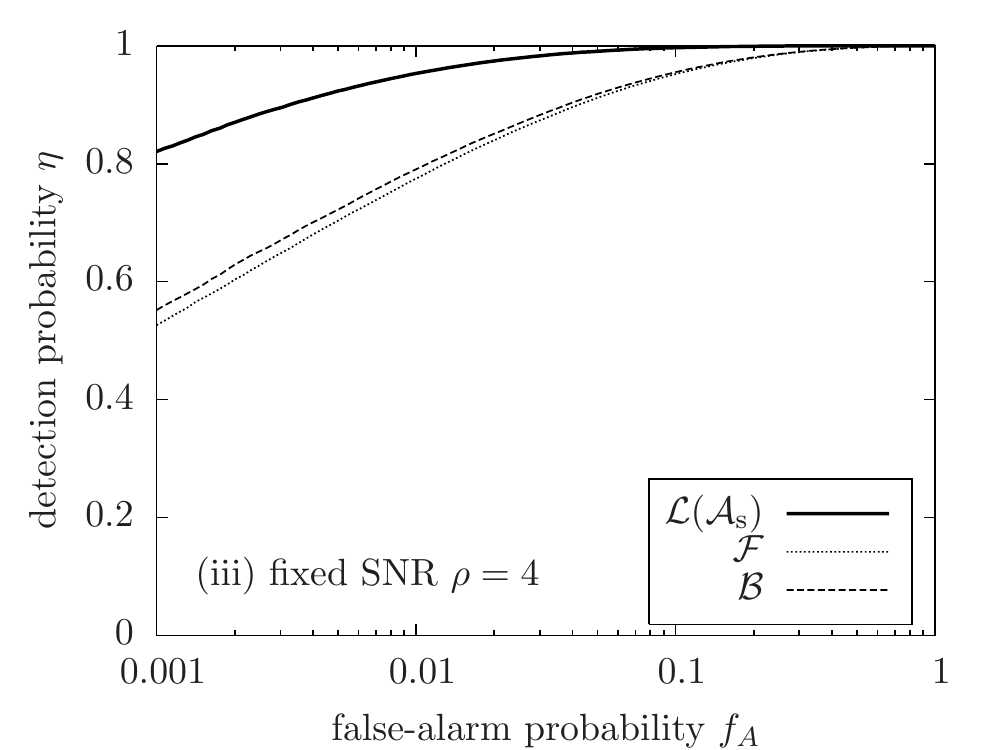}
    \hspace*{-0.8cm}
    \includegraphics[width=0.55\columnwidth,clip]{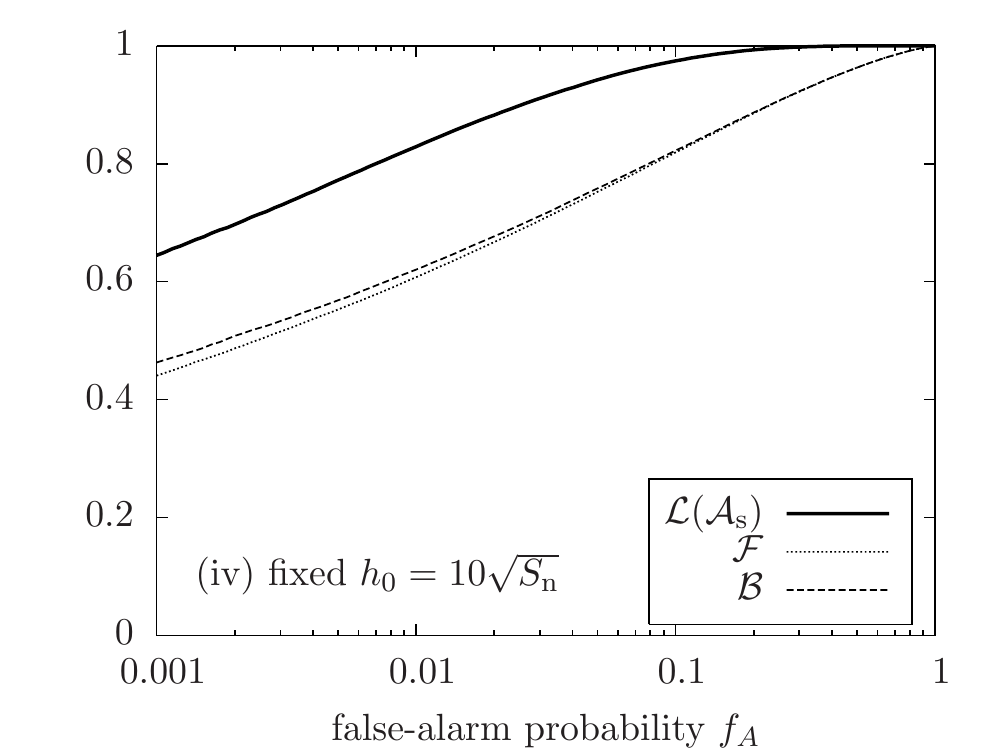}
  }
  \caption{ROC curves $\eta(\fA;\pA)$ comparing $\F$-statistic (\ref{eq:34}), $\B$-statistic
    (\ref{eq:28}), and the perfect-match likelihood ratio $\Lr(\Amp_\sig)$.
    The signal populations $\pA$ consist of randomly distributed $\cosi$ and $\psi$, according to
    the physical prior $\Prior_{\physical}$ (cf.\ section~\ref{sec:what-are-physical}), with (iii) a
    \emph{fixed SNR} of $\rho=4$ and (iv) a \emph{fixed amplitude} of $h_0=10\,\sqrt{\Sn}$.
  }
  \label{fig:roc_example2}
\end{figure}
The results of these simulations are shown in Fig.~\ref{fig:roc_example2}.
We see that in these situations the $\B$-statistic is consistently more powerful than the $\F$-statistic.
This is not surprising, given that the amplitude prior $\Prior_\canonical$ that is implicit in the
$\F$-statistic differs substantially from the injected ``real-world'' isotropic probability distribution
$\pA$ on the orientation of $\vOm$. The $\B$-statistic prior $\Prior_\physical$, on the other hand,
is consistent with $\pA$ by construction (cf.\ section~\ref{sec:relat-betw-canon}). In fact, it can
be argued \cite{2008arXiv0804.1161S} that the Bayes factor with a signal prior that is consistent
with the population injected in the Monte-Carlo simulation is \emph{by construction} optimal in the
sense of the highest expected detection probability at given false-alarm probability. One would
therefore not expect any other detection statistic to outperform $\B(x)$ in the simulations shown in
Fig.~\ref{fig:roc_example2}.

\section{Conclusions}
\label{sec:conclusions}

We have shown that the maximum-likelihood $\F$-statistic can be interpreted as a Bayes factor with a
simple, but unphysical, amplitude prior (and an additional unphysical sky-position weighting).
Using a more physical prior based on an isotropic probability distribution for the unknown
spin-axis orientation of emitting systems, we obtain a new detection statistic, referred to as
$\B$-statistic. Monte-Carlo simulations for signals with random (isotropic) spin-axis
orientations show that the $\B$-statistic more powerful (in terms of its expected detection
probability) than the $\F$-statistic.

The $\F$-statistic is therefore not ``optimal'' in the classical sense.
However, the $\F$-statistic sensitivity appears to be quite comparable to the $\B$-statistic
(see Fig.~\ref{fig:roc_example2}) in the range of parameters considered, while being
computationally more efficient (there are no integrations required), and fully
characterized by a known simple distribution. Interpreting it as a Bayes factor clarifies its role
as a statement about relative probabilities of hypotheses, and allows one to use the $\F$-statistic
within a fully Bayesian framework. Such a choice would be based on the simplicity and computational
efficiency of the $\F$-statistic, despite the fact that it is not ``optimal''.

\ack
  This work has benefited crucially from numerous discussions with colleagues, in
  particular John T. Whelan, Jolien Creighton, Teviet Creighton, Christian
  R\"o{}ver, Holger Pletsch, Graham Woan, Curt Cutler and Chris Messenger.

\section*{References}

\bibliography{biblio}

\providecommand{\newblock}{}
\begin{thebibliography}{10}
\expandafter\ifx\csname url\endcsname\relax
  \def\url#1{{\tt #1}}\fi
\expandafter\ifx\csname urlprefix\endcsname\relax\def\urlprefix{URL }\fi
\providecommand{\eprint}{}

\bibitem{prix06:_cw_review}
Prix R (for the LIGO Scientific Collaboration) 2009 {\em Neutron Stars and
  Pulsars\/} ed Becker W (Springer Verlag) p 651 \texttt{LIGO-P060039-v1},
  (\url{http://dcc.ligo.org/cgi-bin/DocDB/ShowDocument?docid=635})

\bibitem{krolak04:_optim_lisa}
Kr{\'o}lak A, Tinto M and Vallisneri M 2004 {\em Phys.\ Rev.\ D.\/} {\bf 70}
  022003 (\textit{Preprint} \eprint{gr-qc/0401108})

\bibitem{WhelanPrix08:_MLDC1B}
{Whelan} J~T, {Prix} R and {Khurana} D 2008 {\em Class.\ Quant.\ Grav.\/} {\bf
  25} 184029 (\textit{Preprint} \eprint{arXiv:0805.1972[gr-qc]})

\bibitem{2001PhRvD..64d2004P}
{Pai} A, {Dhurandhar} S and {Bose} S 2001 {\em Phys.\ Rev.\ D.\/} {\bf 64}
  042004 (\textit{Preprint} \eprint{gr-qc/0009078})

\bibitem{jks98:_data}
Jaranowski P, Kr{\'o}lak A and Schutz B~F 1998 {\em Phys.\ Rev.\ D.\/} {\bf 58}
  063001

\bibitem{lsc04:_psr_j1939}
Abbott B {\em et~al.\/} (LIGO Scientific Collaboration) 2004 {\em Phys.\ Rev.\
  D.\/} {\bf 69} 082004

\bibitem{lsc06:_coher_scorp_x}
Abbott B {\em et~al.\/} (LIGO Scientific Collaboration) 2007 {\em Phys.\ Rev.\
  D.\/} {\bf 76} 082001 (\textit{Preprint} \eprint{gr-qc/0605028})

\bibitem{2005CQGra..22S1243A}
{Astone} P {\em et~al.\/} 2005 {\em Class.\ Quant.\ Grav.\/} {\bf 22} S1243

\bibitem{PrixWhelan07:_MLDC1}
Prix R and Whelan J~T 2007 {\em Class.\ Quant.\ Grav.\/} {\bf 24} 565

\bibitem{2008PhRvD..78b2001V}
{Veitch} J and {Vecchio} A 2008 {\em Phys.\ Rev.\ D.\/} {\bf 78} 022001
  (\textit{Preprint} \eprint{0801.4313})

\bibitem{2007PhRvD..76d3003C}
{Clark} J, {Heng} I~S, {Pitkin} M and {Woan} G 2007 {\em Phys.\ Rev.\ D.\/}
  {\bf 76} 043003 (\textit{Preprint} \eprint{arXiv:gr-qc/0703138})

\bibitem{2008arXiv0804.1161S}
{Searle} A~C 2008 (\textit{Preprint} \eprint{0804.1161})

\bibitem{2008CQGra..25r4010V}
{Veitch} J and {Vecchio} A 2008 {\em Classical and Quantum Gravity\/} {\bf 25}
  184010 (\textit{Preprint} \eprint{0807.4483})

\bibitem{2008CQGra..25k4038S}
{Searle} A~C, {Sutton} P~J, {Tinto} M and {Woan} G 2008 {\em Class.\ Quant.\
  Grav.\/} {\bf 25} 114038 (\textit{Preprint} \eprint{0712.0196})

\bibitem{2008arXiv0809.2809S}
{Searle} A~C, {Sutton} P~J and {Tinto} M 2008 (\textit{Preprint} \eprint{0809.2809})

\bibitem{2005PhRvD..72j2002D}
{Dupuis} R~J and {Woan} G 2005 {\em Phys.\ Rev.\ D.\/} {\bf 72} 102002

\bibitem{2004CQGra..21S1655U}
{Umst{\"a}tter} R, {Meyer} R, {Dupuis} R~J, {Veitch} J, {Woan} G and
  {Christensen} N 2004 {\em Classical and Quantum Gravity\/} {\bf 21} 1655

\bibitem{2007arXiv0708.3818L}
Abbott B {\em et~al.\/} (LIGO Scientific Collaboration) 2008 {\em Phys.\ Rev.\
  D.\/} {\bf 77} 022001

\bibitem{iraj_thesis_2007}
Gholami I 2007 {\em Data Analysis of Continuous Gravitational Waves\/} Ph.D.
  thesis Albert Einstein Institute

\bibitem{krishnan04:_hough}
{Krishnan} B, {Sintes} A~M, {Papa} M~A, {Schutz} B~F, {Frasca} S and {Palomba}
  C 2004 {\em Phys.\ Rev.\ D.\/} {\bf 70} 082001 (\textit{Preprint}
  \eprint{arXiv:gr-qc/0407001})

\bibitem{bonazzola96:_gravit}
Bonazzola S and Gourgoulhon E 1996 {\em A\&A\/} {\bf 312} 675--690

\bibitem{finn92:_detection}
Finn L~S 1992 {\em Phys.\ Rev.\ D.\/} {\bf 46} 5236--5249

\bibitem{helstrom68:_signal_detection}
Helstrom C~W 1968 {\em Statistical Theory of Signal Detection\/} 2nd ed
  (Oxford: Pergamon Press)

\bibitem{1999kats.book.....S}
{Stuart} A, {Ord} J~K and {Arnold} S 1999 {\em {Kendall's advanced theory of
  statistics. Vol.2A: Classical inference and the linear model}\/} (Arnold)

\bibitem{jaynes:_logic_of_science}
Jaynes E~T 2003 {\em Probability Theory. The Logic of Science\/} (Cambridge
  University Press)

\bibitem{sivia96:_bayesian}
Sivia D~S 2006 {\em Data Analysis. A Bayesian Tutorial\/} (Oxford, University
  Press)

\bibitem{2005blda.book.....G}
{Gregory} P~C 2005 {\em {Bayesian Logical Data Analysis for the Physical
  Sciences: A Comparative Approach with `Mathematica' Support}\/} (Cambridge
  University Press)

\bibitem{conway:Voronoi1984}
Conway J~H and Sloane N~J~A 1984 {\em SIAM Journal on Algebraic and Discrete
  Methods\/} {\bf 5} 294--305
  \urlprefix\url{http://link.aip.org/link/?SML/5/294/1}

\end{thebibliography}

\end{document}